%% file: paper.tex
\PassOptionsToPackage{table}{xcolor}
\PassOptionsToPackage{shortlabels}{enumitem}
\PassOptionsToPackage{nopatch=footnote}{microtype}
\documentclass[sigconf,9pt]{acmart}

\usepackage[english]{babel}
\usepackage{booktabs}
\usepackage{array}
\usepackage{tabularx}
\usepackage{amsmath}
\usepackage{graphicx}
\usepackage{multirow}
\usepackage{xcolor}
\usepackage{xspace}
\usepackage{fontawesome5}
\usepackage{siunitx}
\usepackage{ifthen}
\usepackage{amssymb}         
\usepackage{enumitem}        
\usepackage{cleveref}
\input{theme}                 

\crefname{figure}{Fig.}{Figs.}
\Crefname{figure}{Fig.}{Figs.}
\crefname{equation}{Eq.}{Eqs.}
\Crefname{equation}{Eq.}{Eqs.}
\Crefname{table}{Tab.}{Tabs.}
\crefname{table}{Tab.}{Tabs.}
\crefname{section}{\S}{\S\S}
\Crefname{section}{\S}{\S\S}
\crefname{subsection}{\S}{\S\S}
\Crefname{subsection}{\S}{\S\S}

\newcommand{\skipping}{\textit{weak-skipping}\xspace}
\newcommand{\conditioned}{\textit{weak-conditioned}\xspace}
\newcommand{\adaptive}{\textit{budget-adaptive}\xspace}
\newcommand{\Skipping}{\textit{Weak-skipping}\xspace}
\newcommand{\Conditioned}{\textit{Weak-conditioned}\xspace}
\newcommand{\Adaptive}{\textit{Budget-adaptive}\xspace}

\newcommand{\best}[1]{{\bfseries\boldmath #1}}
\newcommand{\winS}[1]{#1\,\raisebox{0.3ex}{\textcolor{green!55!black}{\tiny$\blacktriangle$}}}
\newcommand{\eqS}[1]{#1\,\raisebox{0.3ex}{\textcolor{black!45}{\tiny$\bullet$}}}

\makeatletter
\newcommand{\change}[3]{%
  \ifthenelse{\equal{#3}{-}}%
    {\def\ch@sz{\scriptsize}\let\ch@wrap\relax\def\ch@fade{!60!white}}%
    {\ifthenelse{\equal{#3}{+}}%
      {\let\ch@sz\relax\let\ch@wrap\textbf\def\ch@fade{}}%
      {\let\ch@sz\relax\let\ch@wrap\relax\def\ch@fade{}}}%
  \ifthenelse{\equal{#2}{+}}%
    {\def\ch@col{green!55!black\ch@fade}\def\ch@sym{$\blacktriangle$}}%
    {\ifthenelse{\equal{#2}{-}}%
      {\def\ch@col{red!70!black\ch@fade}\def\ch@sym{$\blacktriangledown$}}%
      {\def\ch@col{black!45!white\ch@fade}\def\ch@sym{$=$}}}%
  \ifthenelse{\equal{#1}{}}%
    {{\textcolor{\ch@col}{\ch@sz\ch@wrap{\raisebox{0.3ex}{\tiny\ch@sym}}}}}%
    {{\textcolor{\ch@col}{\ch@sz\ch@wrap{\num{#1}\%}\raisebox{0.3ex}{\tiny\ch@sym}}}}%
}
\makeatother

\begin{document}

\acmYear{2026}\copyrightyear{2026}
\setcopyright{cc}
\setcctype[4.0]{by}
\acmConference[SIGCOMM '26]{Workshop on Networks for AI Computing}{August 17--21, 2026}{Denver, CO, USA}
\acmBooktitle{Workshop on Networks for AI Computing (SIGCOMM '26), August 17--21, 2026, Denver, CO, USA}
\acmDOI{10.1145/3789240.3828740}
\acmISBN{979-8-4007-2467-1/26/08}

\title{Budget-Adaptive Routing: Skipping the Weak When the Strong Answers Anyway}

\author{Wei Geng}
\affiliation{%
  \institution{Technical University of Munich}
  \city{Munich}
  \country{Germany}
}
\email{wei.geng@tum.de}

\author{Nitinder Mohan}
\affiliation{%
  \institution{TU Delft}
  \city{Delft}
  \country{Netherlands}
}
\email{n.mohan@tudelft.nl}

\author{Jörg Ott}
\affiliation{%
  \institution{Technical University of Munich}
  \city{Munich}
  \country{Germany}
}
\email{ott@in.tum.de}

\renewcommand{\shortauthors}{Geng et al.}

\begin{abstract}
Edge-cloud inference collaborations are often designed with a routing estimator\footnote{In this paper, in most cases we use \emph{estimator} and \emph{router} interchangeably.} that decides whether to offload each frame from weak models at the edge to stronger models in the cloud. Existing systems place the routing estimator \emph{after} the weak detector, so the weak forward pass still runs even on frames that are later offloaded.
In this paper, we argue that this \conditioned design can be suboptimal when the offload budget varies.
First, we present a competitive \emph{\skipping} estimator ($0.153$\,GFLOPs, ${\sim}29\times$ lighter than the weak detector at $4.49$\,GFLOPs) that extracts routing signal from raw pixels, outperforming the common after-weak placement \conditioned baselines. Second, we show that neither \skipping nor \conditioned placement dominates across the full operating curve, and we propose \adaptive routing, which selects between them by offload budget via two offline-tuned thresholds. 
On PASCAL VOC, our \adaptive router traces the upper accuracy envelope of both fixed placements across the operating range. Our method~\footnote{Artifacts are available at \faGithub~\url{https://github.com/ViGeng/bgt-ada}} reduces per-frame latency by up to $19.1$\,ms (${\sim}30\%$ lower at $\rho{=}0.9$). Besides outperforming SOTA methods, it is surprisingly \textbf{stronger than the strong} model ($+1.7$\,pp over the strong model's peak mAP) at some operating points with far less compute.
\end{abstract}

\input{ccs}


\maketitle

\input{body}
\input{method.tex}
\input{evidence.tex}

\begin{acks}
This work was supported by the Dutch National Growth Fund ``Future Network Services''.
\end{acks}

\newpage
\bibliographystyle{ACM-Reference-Format}
\bibliography{reference}

\input{appendix}

\end{document}

%% file: theme.tex
\input{figstyle/wgpalette}

\colorlet{adapt}{wgTeal}     
\colorlet{skip}{wgAzure}     
\colorlet{cond}{wgVivid}     
\colorlet{skim}{wgBlue}      
\colorlet{edgeml}{wgPink}    
\colorlet{dcsb}{wgRed}       

\colorlet{est}{wgAzure}      
\colorlet{weak}{wgVivid}     
\colorlet{cloud}{wgSlate!50} 
\colorlet{save}{wgGreen}     
\colorlet{penalty}{wgPink}   

\colorlet{benefit}{wgGreen}   
\colorlet{harm}{wgRed}        
\colorlet{neutral}{wgSlate}   
\colorlet{headroom}{wgBlue}   

\colorlet{weakm}{wgAmber}    
\colorlet{estm}{wgBlue}      
\colorlet{strongm}{wgSlate}  
\colorlet{partialm}{wgPurple}
\colorlet{offload}{wgRed}    
\colorlet{keeplocal}{wgGreen}

%% file: figstyle/wgpalette.tex
\definecolor{wgTeal}{HTML}{1C7C6C}\definecolor{wgTealBg}{HTML}{E7F3EF}
\definecolor{wgBlue}{HTML}{2E6FB0}\definecolor{wgBlueBg}{HTML}{E8F1FB}
\definecolor{wgAmber}{HTML}{C9821F}\definecolor{wgAmberBg}{HTML}{FBEEDB}
\definecolor{wgSlate}{HTML}{6E767E}\definecolor{wgSlateBg}{HTML}{EEF1F3}
\definecolor{wgGreen}{HTML}{2E8B57}\definecolor{wgGreenBg}{HTML}{E9F4EC}
\definecolor{wgRed}{HTML}{C4313C}\definecolor{wgRedBg}{HTML}{FBE9EA}
\definecolor{wgPurple}{HTML}{6A51A3}\definecolor{wgPurpleBg}{HTML}{ECE8F4}
\definecolor{wgPink}{HTML}{CC79A7}\definecolor{wgPinkBg}{HTML}{F7EAF1}
\definecolor{wgVivid}{HTML}{D55E00}
\definecolor{wgAzure}{HTML}{0072B2}
\definecolor{wgInk}{HTML}{2B2B2B}
\definecolor{wgEdgeBg}{HTML}{F4F8F7}
\colorlet{wgComment}{wgSlate}

%% file: ccs.tex
\ccsdesc[500]{Networks~Cloud computing}
\ccsdesc[500]{Computer systems organization~n-tier architectures}
\ccsdesc[300]{Computing methodologies~Object detection}
\ccsdesc[500]{Networks~Network performance modeling}

%% file: body.tex
\section{Introduction}
\label{sec:intro}

Edge devices increasingly run visual perception pipelines by offloading computation to the cloud, such as traffic cameras counting vehicles, industrial robots operating on assembly lines, and identity systems checking faces at borders. While some simply stream tasks to the edge or cloud for inference~\cite{geng2026smooth,jiang2018chameleon,li2020reducto,MLSYS2019_6bcfac82,chen2015glimpse}, others pair a \emph{local/edge} \emph{weak} detector that is fast but limited with a \emph{strong} detector in the \emph{cloud} that is more accurate but more costly in latency, compute, bandwidth, and energy. Selective offloading routes each frame so that the cloud complements local inference, reducing overall cost without sacrificing, and often improving, end-to-end accuracy~\cite{cao2023edge,qiu2024optimizing}. This routing decision is constrained by an offload budget that caps the fraction of frames sent to the cloud, given network conditions, compute resources, and application requirements. 

\begin{figure}[tb]
    \centering
    \includegraphics[width=\columnwidth]{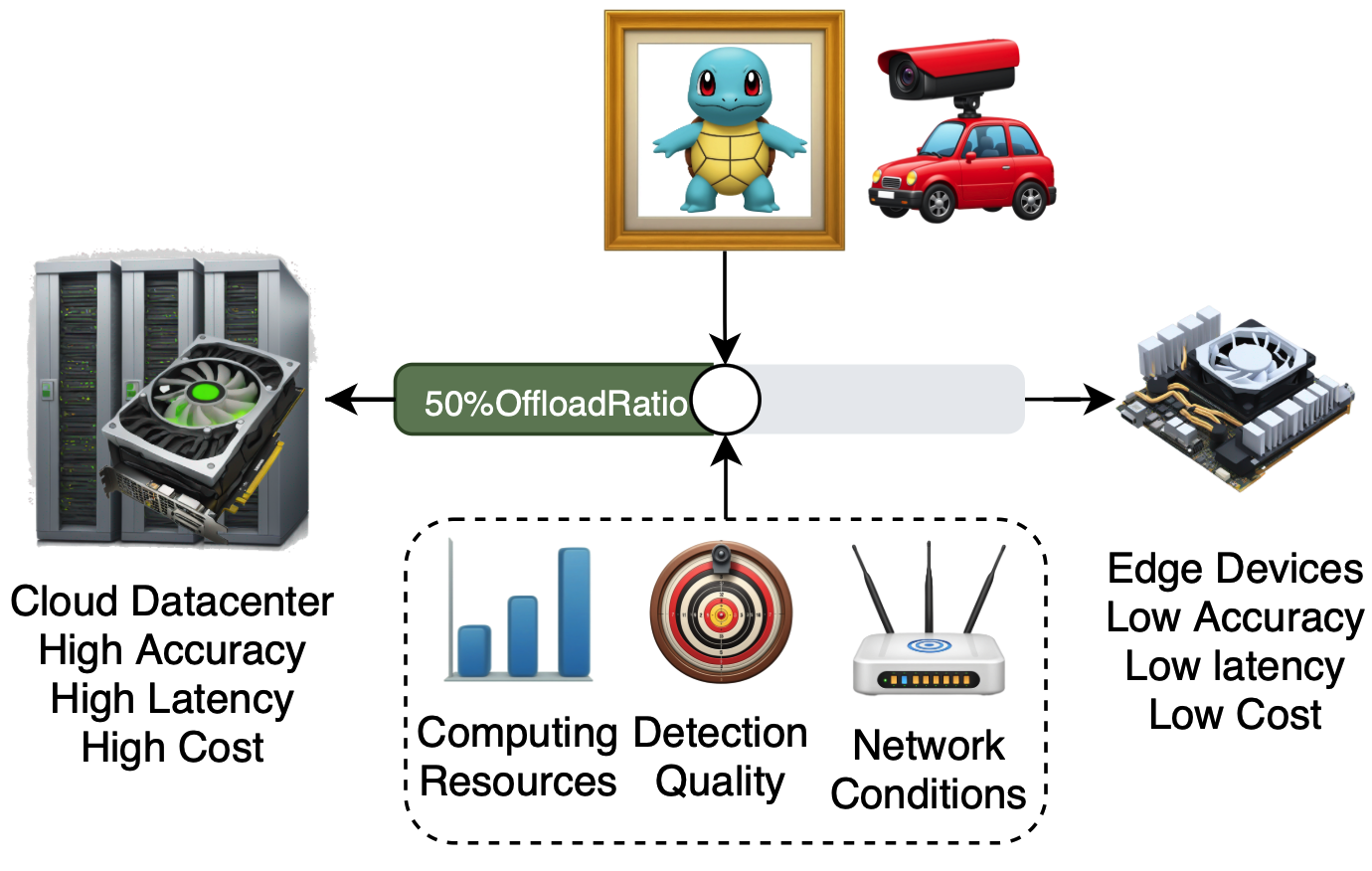}
    \caption{Selective offloading for object detection: a \emph{local} \emph{weak} detector and a \emph{cloud} \emph{strong} detector with a per-frame router under offload budget $\rho$.}
    \label{fig:problem}
\end{figure}

A decade of selective-offloading work~\cite{qiu2024optimizing,cao2023edge,wang2024tiny,teerapittayanon2016branchynet,huang2017multi,kaya2019shallow,geng2026smooth} runs the weak detector on every frame and feeds its output (proposal scores, top-$k$ box statistics, learned embeddings, or intermediate activations) to a \emph{routing estimator} that decides whether to escalate to a stronger detector in the cloud. EdgeML~\cite{qiu2024optimizing} regresses on the top-$25$ proposal features so that uncertain cases can be improved by strong models. DCSB~\cite{cao2023edge} hand-crafts a difficult-case discriminator that directs requests to the strong model at a fixed threshold. Early-exit families such as BranchyNet~\cite{teerapittayanon2016branchynet}, MSDNet~\cite{huang2017multi}, and SDN~\cite{kaya2019shallow} gate \emph{within} the weak network at intermediate layers. Despite their architectural variety, all of these methods share one structural commitment: \emph{the weak forward pass runs on every frame, because the routing decision depends on its output or intermediate features}.

This commitment is misaligned with high-budget deployments. Safety-critical settings such as autonomous driving or security run at a high offload budget, say $\geq 30\%$ of frames sent to the cloud, and there the weak forward pass is wasted on most frames: they are offloaded and answered by the strong model anyway, so paying for the weak pass only adds unnecessary compute and latency.

\begin{figure}[htb]
    \centering
    \includegraphics[width=\columnwidth]{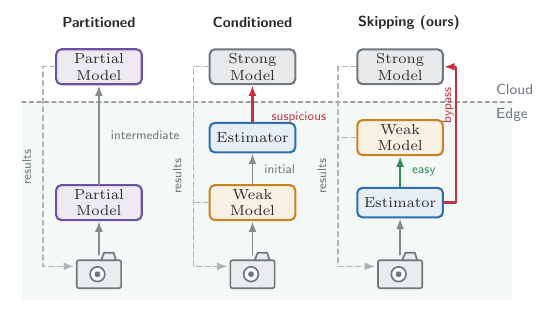}
    \caption{Routing schemes: full partitioned compute~\cite{kang2017neurosurgeon,kaya2019shallow,teerapittayanon2016branchynet,huang2017multi} (left), \conditioned after the weak pass~\cite{cao2023edge,qiu2024optimizing} (middle), and our \skipping (right), which enables \adaptive routing. \textcolor{estm}{Estimator} placement is the axis: it fires after the weak model (middle) or before it (right). \textcolor{keeplocal}{Green} marks the keep-local path, \textcolor{offload}{red} the escalation to the cloud.}
    \label{fig:schemes}
\end{figure}

Structurally, if the estimator is moved \emph{before} the weak model and predicts from the raw image, the serial dependency on the weak model is removed and the weak pass can be skipped for frames that will most likely be offloaded. We call this a \skipping estimator, in contrast to the \conditioned estimators of prior work (shown in \Cref{fig:schemes} left and middle). To our knowledge no previous selective-offloading system has deployed it. The usual assumption is that raw-pixel features are too weak to predict detector failure. Our results challenge this: a $0.15$\,GFLOPs image-only \skipping estimator trained on a binary offload-utility target (\Cref{eq:offloadbin}) is competitive with proposal-feature baselines that require the full $4.49$\,GFLOPs weak forward pass. We attribute this to the assumption that predicting \emph{whether} a frame is difficult is substantially cheaper than predicting \emph{what} it contains.

\Skipping is not, however, strictly better. At low offload budgets the weak forward pass is unavoidable anyway. Since the cloud rarely fires, the \conditioned estimator's richer features come essentially for free. The two placements therefore divide the operating curve into compute regimes: \conditioned and \skipping win in the low and high-budget bands, respectively, and neither dominates the curve.

Taken together, these observations argue for an \emph{adaptive} router that selects between \skipping and \conditioned placements as a function of the offload budget, instead of one fixed placement. We confirm it empirically on PASCAL VOC from compute, latency, and accuracy perspectives.
 This paper contributes:
\begin{enumerate}[(i),itemsep=2pt,topsep=2pt,parsep=0pt,leftmargin=1.6em]
    \item We articulate the \emph{weak-first assumption} latent in the selective-offloading literature, formalize its hidden cost as an \emph{implicit compute tax}. (\Cref{sec:problem}).
    \item We introduce the \emph{first competitive} \skipping estimator for object-detection offloading: a $0.15$\,GFLOPs image-only estimator that matches or exceeds the strongest \conditioned baselines (\Cref{sec:skipping}) detection quality-wise. We also build a lightweight \conditioned estimator (XGBoost on MORIC) that outperforms current \conditioned SOTAs on routing quality at sub-ms inference cost.(\Cref{tab:routing-quality}).
    \item Building on our \skipping and \conditioned estimators, we propose \adaptive routing, which selects between \skipping and \conditioned placements according to the deployment budget. From our simulated study, our approach outperforms SOTAs. (\Cref{sec:adaptive}, \Cref{sec:evidence}).
\end{enumerate}

\section{Problem Statement}
\label{sec:problem}

\subsection{Problem Formulation}
\label{sec:formulation}

Let $\mathcal{I} = (i_1, \dots, i_N)$ be a stream of image frames, $M_w$ a weak local detector with per-frame compute cost $C_w$, and $M_s$ a strong cloud detector with per-frame cost $C_s$ (compute plus network round-trip). For each frame $i_t$ the router produces a binary routing decision $d_t \in \{0, 1\}$ where $d_t = 0$ means ``return $M_w(i_t)$'' and $d_t = 1$ means ``return $M_s(i_t)$''. Given a downstream detection utility $U(\cdot)$ (typically mean Average Precision, mAP), an offload budget $\rho \in (0, 1]$, and an estimator producing a per-frame score $s_t$, the router solves
\begin{equation}
\max_{d_{1:N}}\ \frac{1}{N} \sum_{t=1}^N U\bigl(i_t, M_{d_t}\bigr) \quad \text{s.t.}\quad \frac{1}{N}\sum_{t=1}^N d_t \le \rho.
\label{eq:problem}
\end{equation}

We instantiate $U$ as mAP@0.5, the standard PASCAL VOC metric~\cite{everingham2010pascal}, for comparability with the VOC benchmark and the \conditioned baselines we reproduce. The framework is otherwise metric-agnostic: the proxy reward $\Delta\!\text{AP}$ (\Cref{eq:dap}) can be computed for any $U$, so a stricter COCO-style AP@[.5:.95] is a drop-in substitution we leave to future work.

\subsection{The Implicit Compute Tax}
\label{sec:tax}

In every \conditioned design the weak pass executes on \emph{every} frame, including those answered by the cloud, as an \emph{implicit compute tax}. The expected per-frame compute is
\begin{align}
T_{\text{cond}}(\rho) &= C_w + C_e^{\text{cond}} + \rho \cdot C_s, \label{eq:cond-cost}\\
T_{\text{skip}}(\rho) &= C_e^{\text{skip}} + (1-\rho)\cdot C_w + \rho\cdot C_s, \label{eq:skip-cost}
\end{align}
where $C_e^{\text{cond}}$ and $C_e^{\text{skip}}$ are the estimator costs for the two placements. The tax is $\rho\, C_w - (C_e^{\text{skip}} - C_e^{\text{cond}})$. With our values (MobileNetV3~\cite{howard2019mobilenetv3} vs ResNet50~\cite{he2016deep}, $C_w{=}4.49$, $C_e^{\text{cond}}{\approx}0$, $C_e^{\text{skip}}{=}0.15$, $C_s{=}280.37$\,GFLOPs), the tax reaches $3.90$\,GFLOPs at $\rho{=}0.9$, approaching the full weak-detector cost. In addition, a \conditioned router cannot start until the weak forward pass finishes ($24.70$\,ms here), so $C_w$ is a serial wall-clock dependency on every frame.

The tax reflects a fallback mindset: the cloud as a backstop for the weak detector. Beside the compute and latency costs, it may also increase the jitter of the system (we leave further analysis to a follow-up work), which is undesirable for real-time applications. We don't have to pay the tax if we treat the cloud as a collaborator that complements local inference instead of a fallback, illustrated as a branch topology in \Cref{fig:schemes}-right. Then the practical question is \emph{whether an image-only lightweight estimator can catch enough routing signal without paying the tax?}

%% file: method.tex
\section{Method}
\label{sec:method}

\subsection{\Skipping Estimator}
\label{sec:skipping}

\begin{figure}[t]
\centering
\includegraphics[width=0.93\columnwidth]{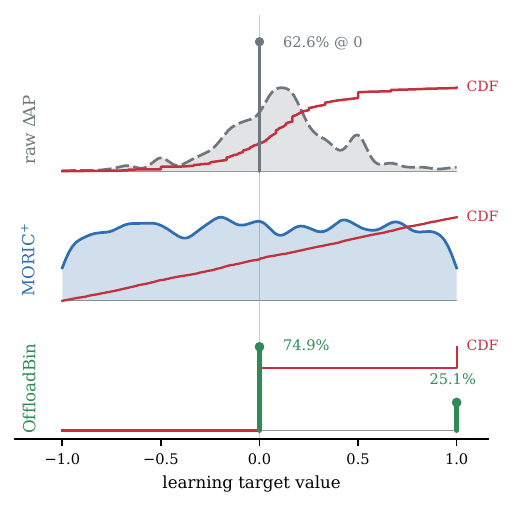}
\caption{Per-frame learning targets on VOC test ($N{=}3105$). Top: raw $\Delta\!\text{AP}$ is degenerate, with $62.6\%$ of frames exactly at $0$ (stem) and short signed tails ($25.1\%{>}0$, $12.3\%{<}0$). Middle: MORIC$^{+}$ spreads this into a smooth, symmetric regression target on $[-1,1]$. Bottom: OffloadBin ($\mathbf{1}[\Delta\!\text{AP}{>}0]$) is the binary classification target, a ${\sim}1{:}3$ split ($25.1\%$ positive).}
\label{fig:target-dist}
\end{figure}

\textbf{Proxy metric.}
A \skipping estimator can only out-cheap a \conditioned one if it learns to predict \emph{whether offloading helps}, not the contents of the frame. A naive per-frame target like $\Delta\!\text{AP}$ ($= \text{cloud AP} - \text{local AP}$), is misleading because detection AP is computed dataset-wide via a single global Precision-Recall (PR) curve, so a single frame's contribution depends on every other frame. Inspired by EdgeML~\cite{qiu2024optimizing}, we use the \emph{contextual offloading reward}: for frame $i_t$, the per-frame reward
\begin{equation}
\Delta\!\text{AP}(i_t) \;=\; \text{AP}\bigl(\text{swap}_t \to s\bigr) \;-\; \text{AP}\bigl(\text{all-weak}\bigr),
\label{eq:dap}
\end{equation}
i.e., the change in dataset-wide AP@0.5 obtained by replacing the local detections on $i_t$ with the cloud detections while holding all other frames at their local outputs. Computing it offline over the training set is $\mathcal{O}(N)$ via a precomputed-IoU merge swap, and the resulting reward depends only on the dataset, not on the system at inference time.
\Cref{eq:dap} produces a long-tailed signed signal that is hard to regress directly, as shown by the raw $\Delta\!\text{AP}$ ridge (top) of~\Cref{fig:target-dist}. We extract two estimator-friendly targets:
\begin{equation}
\text{MORIC}^{+}(i_t) =
\begin{cases}
F^{+}_{\Delta\text{AP}}\bigl(\Delta\!\text{AP}(i_t)\bigr)        & \Delta\!\text{AP}(i_t) > 0, \\
0                                                                & \Delta\!\text{AP}(i_t) = 0, \\
F^{-}_{\Delta\text{AP}}\bigl(\Delta\!\text{AP}(i_t)\bigr) - 1    & \Delta\!\text{AP}(i_t) < 0,
\end{cases}
\label{eq:moricplus}
\end{equation}
\begin{equation}
\text{OffloadBin}(i_t) = \mathbf{1}\!\bigl[\Delta\!\text{AP}(i_t) > 0\bigr],
\label{eq:offloadbin}
\end{equation}
where $F^{+}$ and $F^{-}$ are the empirical CDFs of $\Delta\!\text{AP}$ over the frames where offloading strictly helps ($\Delta\!\text{AP}{>}0$) and strictly hurts ($\Delta\!\text{AP}{<}0$), respectively; $\text{MORIC}^{+}(i_t)\in[-1,1]$ and $\text{OffloadBin}(i_t)\in\{0,1\}$. MORIC$^{+}$ generalises EdgeML's MORIC~\cite{qiu2024optimizing} by splitting the CDF at zero, which equalises positive and negative magnitudes and gives a symmetric loss landscape around the routing boundary. OffloadBin instead reduces the problem to a binary class label, which we train with focal loss~\cite{lin2017focal} to handle the positive-class imbalance. On VOC the raw $\Delta\!\text{AP}$ is long-tailed with $62.6\%$ of frames at exactly zero, $25.1\%$ positive, and $12.3\%$ negative (\Cref{fig:target-dist}), so direct regression wastes capacity on the zero spike. In other words, the estimator is always lazily and blindly predicting the majority class zero, which already yields good rewards. As a result, it fails to learn the routing boundary. Both transformed targets decouple hardness from content, which is what lets raw pixels suffice. OffloadBin gives our best routing quality and MORIC$^{+}$ is the softer-signal variant reported in the per-ratio sweep.

\textbf{Architecture and calibration.}
A \skipping estimator $f_{\text{skip}}: I \mapsto \hat{s}_{\text{skip}}$ produces a routing score from the raw image alone, where $\hat{s}_{\text{skip}}\in[0,1]$ estimates $P(\Delta\!\text{AP}{>}0)$ when trained on OffloadBin (or the predicted $\text{MORIC}^{+}\in[-1,1]$); a higher score means offloading is more likely to help (not how much to help), instantiating the generic per-frame score $s_t$ of \Cref{sec:formulation}. Our reference instantiation is a highly compressed MobileNetV2-Lite~\cite{sandler2018mobilenetv2} backbone ($128{\times}128$ input, $0.15$\,GFLOPs, $0.54$\,M parameters) trained on OffloadBin, though the framework is agnostic to the backbone and proxy. A thresholder $\pi_\rho$~\cite{guo2017calibration} then maps the score stream to binary decisions $d_t{=}\pi_\rho(\hat{s}_{\text{skip}})$ so that the offloaded fraction matches the requested budget $\rho$ without test-set lookahead. Concretely, $\pi_\rho$ keeps a running estimate of the $(1{-}\rho)$ quantile of the incoming scores and offloads any frame scoring above it, nudging the threshold as the stream drifts so the realized offload rate stays near $\rho$. Ratio control thus stays orthogonal to the score: any estimator emitting a calibrated per-frame score drops into the same thresholder. We leave a full treatment of the thresholder, including drift handling, finite-window error, and the resulting jitter, to follow-up work.

\subsection{\Adaptive Routing}
\label{sec:adaptive}

The accuracy side of the placement mirrors the compute side of \Cref{sec:tax}. At low budgets, \skipping wastes richer signals from the weak forward pass on most frames because most of them comes for free. At high budgets, \conditioned pays the tax on most frames because they will be offloaded anyway. Every fixed-placement router is therefore suboptimal on part of the operating curve.

\textbf{Formulation.} A \adaptive router holds two estimators, $f_{\text{skip}}\!:\!I\mapsto\hat{s}_{\text{skip}}$ and $f_{\text{cond}}\!:\!(I,M_w(I))\mapsto\hat{s}_{\text{cond}}$ ($\hat{s}_{\text{cond}}\in[0,1]$, read from the image plus weak output), plus a binary arbiter $\alpha(\rho)\in\{0,1\}$ and the same thresholder $\pi_\rho$ (\Cref{sec:skipping}), now shared across both estimators. The per-frame decision is
\begin{equation}
d_t \;=\; \pi_\rho\!\left(\alpha(\rho)\,\hat{s}_{\text{skip}}(i_t) + (1{-}\alpha(\rho))\,\hat{s}_{\text{cond}}(i_t,M_w(i_t))\right).
\label{eq:adaptive-score}
\end{equation}
\begin{equation}
    \alpha(\rho)\;=\;\arg\max_{a\in\{0,1\}}\;\widehat{\mathrm{AP}}\bigl(\rho,\;a\,f_{\text{skip}}+(1{-}a)\,f_{\text{cond}}\bigr),
\label{eq:alpha}
\end{equation}
where $\widehat{\mathrm{AP}}$ is offline AP@0.5 on a held-out tuning split. Evaluated per budget, \Cref{eq:alpha} makes $\alpha(\rho)$ piecewise constant with two crossovers, splitting the operating range into three regimes:
\begin{equation}
\alpha(\rho)=
\begin{cases}
0 & \rho < \rho_{\text{frontier}}\quad(\conditioned),\\[2pt]
1 & \rho_{\text{frontier}} \le \rho < \rho_{\text{ceiling}}\quad(\skipping),\\[2pt]
0 & \rho \ge \rho_{\text{ceiling}}\quad(\conditioned).
\end{cases}
\label{eq:regimes}
\end{equation}
The two thresholds are the budgets at which the winning placement flips on the tuning split. At the low crossover $\rho_{\text{frontier}}$ the weak pass on offloaded frames turns from a near-free byproduct into pure tax, so \skipping starts to win. At the high crossover $\rho_{\text{ceiling}}$ the fewer frames still kept local carry enough weight that the richer \conditioned signal wins again. We fit both once, offline, by sweeping $\rho$ and reading off the two switch points (on VOC, $\rho_{\text{frontier}}{=}0.3$ and $\rho_{\text{ceiling}}{=}0.8$). At runtime the arbiter is a constant-time lookup on $\rho$.

%% file: evidence.tex
\section{Preliminary Evidence}
\label{sec:evidence}

We report preliminary results on PASCAL VOC~\cite{everingham2010pascal} based on the setup in \Cref{tab:setup}: the weak/strong mAP gap is small which makes routing genuinely difficult and any larger gap can make the benefits \emph{more pronounced}. Three claims are verified:
\begin{enumerate}[(i),itemsep=2pt,topsep=2pt,parsep=0pt,leftmargin=1.6em]
    \item \Skipping estimators are competitive with \conditioned baselines on routing quality, and our \conditioned estimator outperforms current SOTAs (\Cref{subsec:routing-quality}, \Cref{tab:routing-quality,tab:map-by-rho}, \Cref{fig:map-vs-rho}).
    \item The compute advantage of \skipping routing at high offload budget is real GFLOPs- and latency-wise. (\Cref{subsec:compute-advantage}, \Cref{fig:compute}).
    \item A \adaptive router that selects between the two placements traces the lowest cost across the operating curve and Pareto-dominates either fixed placement on the joint accuracy-compute frontier, outperforming all existing baselines (\Cref{subsec:adaptive-dominates}, \Cref{fig:map-vs-rho,tab:map-by-rho}).
\end{enumerate}

\begin{table}[h]
\centering
\footnotesize
\caption{Experiment setup (PASCAL VOC).}
\label{tab:setup}
\setlength{\tabcolsep}{4pt}
\renewcommand{\arraystretch}{1.05}
\begin{tabular}{@{}>{\raggedright\arraybackslash}p{0.17\linewidth}
                  >{\raggedright\arraybackslash}p{0.79\linewidth}@{}}
\toprule
\multicolumn{2}{@{}l}{\textit{Detectors} \scriptsize(profiled on Nvidia~A40)} \\
Weak $M_w$    & \path{fasterrcnn_mobilenet_v3_large_fpn}~\cite{ren2015fasterrcnn,howard2019mobilenetv3,lin2017feature}; $C_w{=}4.49$\,GFLOPs / $24.70$\,ms; mAP\,$=0.760$ \\
Strong $M_s$  & \path{fasterrcnn_resnet50_fpn_v2}~\cite{ren2015fasterrcnn,he2016deep,lin2017feature}; $C_s{=}280.37$\,GFLOPs / $44.12$\,ms; mAP\,$=0.791$ \\
\midrule
\multicolumn{2}{@{}l}{\textit{Estimators} \scriptsize(ours, \Cref{sec:method})} \\
\Skipping     & MobileNetV2-Lite~\cite{sandler2018mobilenetv2} on OffloadBin or MORIC$^{\!+}$ \\
\Conditioned  & XGBoost~\cite{chen2016xgboost} on MORIC \\
\Adaptive     & offline-tuned $\rho_{\text{frontier}}$, $\rho_{\text{ceiling}}$ \\
\midrule
\multicolumn{2}{@{}l}{\textit{Prior \conditioned baselines}} \\
EdgeML~\cite{qiu2024optimizing} & proposal-level, after weak detector \\
DCSB~\cite{cao2023edge}         & fixed-rule, after weak detector \\
\midrule
\multicolumn{2}{@{}l}{\textit{Reference points}} \\
Trivial   & always-weak, always-strong, uniformly random \\
Oracle    & per-frame $\Delta\!\text{AP}$ ground-truth gain \\
\bottomrule
\end{tabular}
\end{table}

\subsection{\Skipping is Empirically Viable}
\label{subsec:routing-quality}

\Cref{tab:routing-quality} reports an overview across all estimators on VOC Test. Our \skipping MobileNetV2-Lite estimator trained on \textit{OffloadBin} attains a Spearman rank correlation of $0.557$ with the per-frame oracle gain, peak end-to-end mAP@0.5 of $0.808$, and AUC$_\rho$\,=\,$0.795$ over the offload-budget sweep. It \emph{outperforms} the strongest \conditioned baseline we could construct (XGBoost on MORIC: $0.472$ / $0.804$ / $0.794$) on every metric, despite running before the weak detector and never seeing its features. Both EdgeML~\cite{qiu2024optimizing} and DCSB~\cite{cao2023edge} sit well below either family. The MORIC$^{\!+}$ regression target lands between the two families and gives a useful soft signal for the smaller-trunk variant.

\begin{table}[t]
\centering
\footnotesize
\caption{Routing quality on VOC (single-pass, seed $42$, mAP at IoU\,$0.5$). Spearman $\rho_s$ vs.\ the per-frame oracle; Peak~mAP and AUC$_\rho$ (area under the mAP--$\rho$ curve) over $\rho\in[0,1]$ with step $0.1$. \textcolor{green!45!black}{Green deltas} on Peak~mAP are relative \% over always-weak ($0.760$). \conditioned methods additionally pay $C_w{=}4.49$\,GFLOPs upstream. ``--'' = no continuous score.}
\label{tab:routing-quality}
\setlength{\tabcolsep}{3.5pt}
\renewcommand{\arraystretch}{1.05}
\begin{tabular}{@{}lcccc@{}}
\toprule
Estimator & $\rho_s$ & Peak mAP & AUC$_\rho$ & GFLOPs \\
\midrule
\multicolumn{5}{@{}l}{\textit{Reference points}} \\
\quad Always weak                                & $0.000$ & $0.760$~\change{0}{0}{-} & $0.760$ & $0$ \\
\quad Always strong                              & $0.000$ & $0.791$~\change{4.08}{+}{-} & $0.791$ & $0$ \\
\quad Uniform random                             & $0.000$ & -- & $0.777$ & $0$ \\
\quad Oracle ($\Delta\!\text{AP}$)               & --      & $0.827$~\change{8.82}{+}{-} & $0.816$ & --  \\
\midrule
\multicolumn{5}{@{}l}{\textit{\Conditioned} \scriptsize(after weak)} \\
\quad EdgeML~\cite{qiu2024optimizing}            & $0.138$ & $0.793$~\change{4.34}{+}{-} & $0.784$ & $\approx 0$ \\
\quad DCSB~\cite{cao2023edge}                    & $0.359$ & $0.789$~\change{3.82}{+}{-} & --      & $\approx 0$ \\
\quad XGBoost~\cite{chen2016xgboost} / MORIC \emph{(Ours)}              & $0.472$ & $0.804$~\change{5.79}{+}{-} & $0.794$ & $\approx 0$ \\
\cmidrule(l){1-5}
\multicolumn{5}{@{}l}{\textit{\Skipping} \scriptsize(before weak)} \\
\quad MV2-Lite + MORIC$^{\!+}$ \emph{(Ours)}     & $0.350$ & $0.801$~\change{5.39}{+}{-} & $0.790$ & $0.15$ \\
\quad MV2-Lite + OffloadBin \emph{(Ours)}        & \best{$0.557$} & \best{$0.808$}~\change{6.32}{+}{-} & \best{$0.795$} & $0.15$ \\
\cmidrule(l){1-5}
\multicolumn{5}{@{}l}{\textit{\Adaptive} \scriptsize($\rho_{\text{frontier}}{=}0.3$, $\rho_{\text{ceiling}}{=}0.8$)} \\
\quad OffloadBin\,$\leftrightarrow$\,MORIC \emph{(Ours)} & --$^{\ddagger}$ & \best{$0.808$}~\change{6.32}{+}{-} & \best{$0.796$} & $0.15^{\dagger}$ \\
\bottomrule
\end{tabular}
\flushleft\scriptsize $^{\dagger}$ Adaptive uses $f_{\text{cond}}$ for $\rho{\le}0.2$ and $\rho{\ge}0.8$, $f_{\text{skip}}$ for $0.3{\le}\rho{\le}0.7$ (offline arbitration, \Cref{eq:alpha}). Reported GFLOPs are the skipping-branch cost; the \conditioned branch additionally pays $C_w$. $^{\ddagger}$ Spearman is undefined for an arbiter that selects a different score per $\rho$; both component scores' $\rho_s$ are reported above.
\end{table}

These results indicate that raw images are sufficient for object-detection routing. With $0.15$\,GFLOPs of estimator compute, we obtain a routing signal that exceeds a strong proposal-feature baseline whose signal depends on the weak model's $4.49$\,GFLOPs. This confirms the suspicion raised in \Cref{sec:intro}: predicting \emph{whether} a frame is difficult is substantially cheaper than predicting \emph{what} it contains.

\subsection{The Compute Advantage is Material}
\label{subsec:compute-advantage}

\begin{figure*}[t]
\centering
\includegraphics[width=\linewidth]{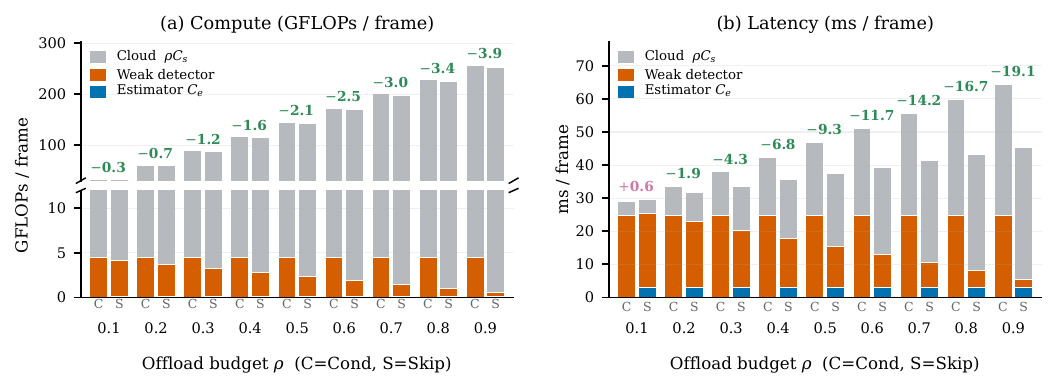}
\caption{Expected per-frame cost vs.\ offload budget $\rho$ on VOC. Each bar stacks estimator $C_e$ (\textcolor{est}{blue}), weak detector (\textcolor{weak}{orange}), and cloud $\rho C_s$ (\textcolor{cloud}{gray}). Paired bars are \textbf{C} (\conditioned, weak on every frame) and \textbf{S} (\skipping, weak on only the $(1{-}\rho)$ kept frames); each pair's label is the \textcolor{save}{cond$-$skip saving} (\textcolor{penalty}{pink} when \skipping costs more). \textbf{(a)}~Compute (GFLOPs, broken $y$-axis; $C_e$ too small to see). \textbf{(b)}~Latency (ms), which adds the serial weak-pass dependency.}
\label{fig:compute}
\end{figure*}

\Cref{fig:compute} decomposes the expected per-frame cost for \conditioned (C) and \skipping (S) at each budget $\rho$ into three stacked components: estimator $C_e$ (\textcolor{est}{blue}), weak detector (\textcolor{weak}{orange}), and cloud $\rho C_s$ (\textcolor{cloud}{gray}). The cloud band dominates both placements equally ($\rho C_s$), so the saving comes entirely from the device side: in the S bars the \textcolor{weak}{orange} weak-detector band shrinks with $(1{-}\rho)C_w$ instead of the constant $C_w$ paid in C, at the cost of adding a thin \textcolor{est}{blue} estimator band ($C_e{=}0.15$\,GFLOPs).

\textbf{Compute (panel a).} The GFLOP saving is monotone in $\rho$: from $0.3$\,GFLOPs at $\rho{=}0.1$ through $1.6$ at $\rho{=}0.4$ and $2.5$ at $\rho{=}0.6$, to $3.9$ at $\rho{=}0.9$, approaching one full weak pass ($4.49$\,GFLOPs) as $\rho{\to}1$. The breakeven is very low, $\rho^\ast{=}C_e/C_w{\approx}0.03$: \skipping is cheaper for essentially all operating budgets, since the estimator pays for itself once it avoids more than one weak pass in ${\sim}30$ frames.

\textbf{Latency (panel b).} The wall-clock picture differs because it captures a \emph{serial} dependency. A \conditioned router cannot produce a score until the weak forward pass finishes ($24.70$\,ms), whereas a \skipping router runs $f_{\text{skip}}$ ($3.08$\,ms) in its place on offloaded frames. At $\rho{=}0.1$ \skipping is $0.6$\,ms \emph{slower} (\textcolor{penalty}{pink} label) because the estimator overhead slightly exceeds the small saving from skipping only $10\%$ of weak passes. The crossover sits at $\rho^\ast_{\text{ms}}{\approx}0.12$. Beyond it the saving grows near-linearly, up to ${\sim}30\%$ end-to-end latency reduction. For latency-critical deployments with higher $\rho$
, this saving alone justifies the \skipping placement.

\begin{table*}[t]
\centering
\small
\caption{End-to-end mAP@0.5 on VOC across offload budgets $\rho\in\{0.1,\dots,0.9\}$ (single seed). Each cell is the accuracy at that $\rho$. \winS{}~beats and \eqS{}~matches the always-strong model ($0.791$), an unmarked cell is below it. \textbf{Bold} is the column best. On the \adaptive row a superscript marks the selected branch (S\,$=$\,\skipping, C\,$=$\,\conditioned).}
\label{tab:map-by-rho}
\setlength{\tabcolsep}{4pt}
\resizebox{\linewidth}{!}{%
\begin{tabular}{@{}lccccccccc@{}}
\toprule
$\rho$ & $0.1$ & $0.2$ & $0.3$ & $0.4$ & $0.5$ & $0.6$ & $0.7$ & $0.8$ & $0.9$ \\
\midrule
\multicolumn{10}{@{}l}{\textit{Reference}} \\
\quad Weak only (constant)                  & $0.760$ & $0.760$ & $0.760$ & $0.760$ & $0.760$ & $0.760$ & $0.760$ & $0.760$ & $0.760$ \\
\quad Strong only (constant)                & $0.791$ & $0.791$ & $0.791$ & $0.791$ & $0.791$ & $0.791$ & $0.791$ & $0.791$ & $0.791$ \\
\quad Random offloader                      & $0.765$ & $0.769$ & $0.772$ & $0.776$ & $0.778$ & $0.781$ & $0.784$ & $0.786$ & $0.788$ \\
\quad Oracle ($\Delta\!\text{AP}$)          & \winS{$0.799$} & \winS{$0.815$} & \winS{$0.823$} & \winS{$0.827$} & \winS{$0.827$} & \winS{$0.827$} & \winS{$0.826$} & \winS{$0.824$} & \winS{$0.817$} \\
\midrule
\multicolumn{10}{@{}l}{\textit{\Conditioned}} \\
\quad EdgeML~\cite{qiu2024optimizing}       & $0.760$ & $0.765$ & $0.789$ & \winS{$0.793$} & \eqS{$0.791$} & \eqS{$0.791$} & \eqS{$0.791$} & \eqS{$0.791$} & \eqS{$0.791$} \\
\quad XGBoost on MORIC \emph{(Ours)}        & \best{$0.776$} & \best{$0.786$} & \winS{$0.794$} & \winS{$0.799$} & \winS{$0.803$} & \winS{$0.804$} & \winS{$0.803$} & \best{\winS{$0.803$}} & \best{\winS{$0.798$}} \\
\midrule
\multicolumn{10}{@{}l}{\textit{\Skipping}} \\
\quad MV2 + MORIC$^{\!+}$ \emph{(Ours)}     & $0.771$ & $0.778$ & $0.785$ & \eqS{$0.791$} & \winS{$0.796$} & \winS{$0.800$} & \winS{$0.801$} & \winS{$0.801$} & \best{\winS{$0.798$}} \\
\quad MV2 + OffloadBin \emph{(Ours)}        & $0.773$ & $0.785$ & \best{\winS{$0.799$}} & \best{\winS{$0.806$}} & \best{\winS{$0.808$}} & \best{\winS{$0.805$}} & \best{\winS{$0.804$}} & \winS{$0.799$} & \winS{$0.794$} \\
\midrule
\multicolumn{10}{@{}l}{\textit{\Adaptive (regime in superscript)}} \\
\quad MV2/OffloadBin\,$\leftrightarrow$\,XGB/MORIC \emph{(Ours)} & \best{$0.776^{\text{C}}$} & \best{$0.786^{\text{C}}$} & \best{\winS{$0.799^{\text{S}}$}} & \best{\winS{$0.806^{\text{S}}$}} & \best{\winS{$0.808^{\text{S}}$}} & \best{\winS{$0.805^{\text{S}}$}} & \best{\winS{$0.804^{\text{S}}$}} & \best{\winS{$0.803^{\text{C}}$}} & \best{\winS{$0.798^{\text{C}}$}} \\
\bottomrule
\end{tabular}}
\end{table*}

\begin{figure}[t]
\centering
\includegraphics[width=\columnwidth]{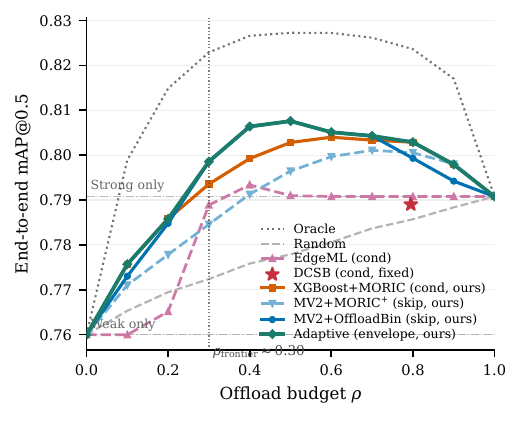}
\caption{End-to-end mAP@0.5 vs.\ offload budget $\rho$ on VOC. \adaptive (\textcolor{adapt}{teal}) is the pointwise upper hull of \skipping (\textcolor{skip}{blue}) and \conditioned (\textcolor{cond}{orange}), with crossovers at $\rho_{\text{frontier}}{\approx}0.3$, $\rho_{\text{ceiling}}{\approx}0.8$.}
\label{fig:map-vs-rho}
\end{figure}

\subsection{\Adaptive Achieves Upper Envelope}
\label{subsec:adaptive-dominates}


\emph{(1) Neither fixed placement dominates the operating curve.} \Cref{fig:map-vs-rho} shows \conditioned (XGBoost on MORIC) mAP leads at low and very high budgets ($\rho{\le}0.2$ and $\rho{\ge}0.8$, by $0.1$--$0.4$\,pp), while \skipping (MV2-Lite + OffloadBin) dominates in the mid-budget regime ($\rho{\in}\{0.3,\dots,0.7\}$). \Cref{tab:map-by-rho} shows that two crossovers bracket the skipping band: at $\rho_{\text{frontier}}{\approx}0.3$ the weak forward pass on offloaded frames becomes a pure tax and skipping it yields both compute savings and better frame selection; at $\rho_{\text{ceiling}}{\approx}0.8$ the few remaining local frames carry enough weight that the richer \conditioned signal again dominates the per-frame compute saved by skipping, also shown in \Cref{fig:map-vs-rho} frame selection quality-wise.

\emph{(2) \Adaptive traces the upper envelope of both placements and strictly dominates all baselines.} A two-threshold offline arbiter ($\rho_{\text{frontier}}{=}0.3$, $\rho_{\text{ceiling}}{=}0.8$) produces the per-budget maximum (last row of \Cref{tab:map-by-rho}, \textcolor{adapt}{teal} curve in \Cref{fig:map-vs-rho}): it picks \conditioned at $\rho{\le}0.2$ and $\rho{\ge}0.8$, \skipping at $0.3{\le}\rho{\le}0.7$. This envelope achieves the \textbf{highest} peak mAP@0.5 of any router, $0.808$, which is $+1.7$\,pp over the strong model itself, and the highest AUC$_\rho{=}0.796$ (\Cref{tab:routing-quality}). Compared with prior \conditioned SOTAs, the \adaptive system leads EdgeML~\cite{qiu2024optimizing} by $0.7$--$2.1$\,pp across the full budget range (the strong model itself leads the weak model by only $3.1$\,pp). The widest gap is at $\rho{=}0.2$ ($0.786$ vs.\ $0.765$) and the narrowest is at $\rho{=}0.9$ ($0.798$ vs.\ $0.791$) where EdgeML's native threshold saturates to always-offload. DCSB~\cite{cao2023edge} is a fixed binary rule locked to a single operating point ($\rho{\approx}0.79$, mAP$=0.789$). At the same budget the \adaptive router reaches $0.803$, and its peak ($0.808$) exceeds DCSB by $1.9$\,pp while offering continuous budget tunability. The envelope sits $1.9 \sim 2.9$\,pp below the offline oracle, bounding what routing quality alone can recover.

\emph{(3) The dominance extends to the joint accuracy-compute frontier inside the skipping band.} For $0.3{\le}\rho{\le}0.7$, \adaptive inherits \skipping's compute profile: $2.1$\,GFLOPs/frame less than any \conditioned method at $\rho{=}0.5$ and $3.0$\,GFLOPs/frame less at $\rho{=}0.7$ (the top of the band; \Cref{fig:compute}a), with wall-clock savings reaching $14.2$\,ms (${\sim}26\%$ reduction) at $\rho{=}0.7$ (\Cref{fig:compute}b). At $\rho{<}\rho_{\text{frontier}}$ the weak pass runs on nearly all frames regardless, so switching to \conditioned costs no incremental compute. At $\rho{\ge}\rho_{\text{ceiling}}$ \adaptive has the option to trade the skipping compute saving for the $0.4$\,pp accuracy lift of \conditioned. Inside the skipping band the \adaptive router is therefore never worse in accuracy \emph{and} never worse in compute than either fixed placement alone.

\section{Related Work}
\label{sec:related}

Selective-offloading admits a clean taxonomy along one architectural axis: \emph{when does the estimator fire, relative to the weak model?}

\textbf{\Conditioned estimators.} Shown in \Cref{fig:schemes}-middle, the estimator fires \emph{after} the weak model and consumes its outputs. EdgeML~\cite{qiu2024optimizing} regresses on the top-$25$ proposal features and DCSB~\cite{cao2023edge} hand-crafts a difficult-case discriminator. Confidence thresholds and learned detection embeddings~\cite{wang2024tiny} also belong to this class. They exploit rich signals but pay \emph{implicit compute tax} unconditionally.

\textbf{Mid-network gates and early exit.} This is essentially \conditioned estimators. BranchyNet~\cite{teerapittayanon2016branchynet}, MSDNet~\cite{huang2017multi}, and SDN~\cite{kaya2019shallow} fuse the estimator into the weak network and gate at intermediate layers. They reduce average weak-model cost on easy frames but do not skip the "Early Pass" entirely. Neurosurgeon~\cite{kang2017neurosurgeon} and SPINN~\cite{laskaridis2020spinn,li2018edge,matsubara2022split} partition or progressively split a single network across device and cloud, which is related but distinct.

\textbf{Cascaded inference.} Classical cascades~\cite{viola2001rapid} and modern cascade detectors~\cite{cai2018cascade} escalate on uncertainty. As detailed in \Cref{sec:adaptive}, they use the cheap stage to \emph{produce predictions} whereas \adaptive routing uses it to \emph{route}.

\textbf{\Skipping estimators.} The estimator fires \emph{before} the weak model from the raw image. Early image-complexity heuristics and learned image-level routers in our work belong to this class. They allow the weak pass to be skipped entirely on offloaded frames but have often been considered too weak for detection routing, a premise challenged by our results.

\section{Conclusion and Future Work}
\label{sec:conclusion}

Estimator placement is a design axis the selective-offloading literature has implicitly fixed. We exhibited both a lightweight image-only \skipping estimator and a \conditioned estimator that outperform the SOTAs that we could reproduce from the literature, on both routing quality and compute. We showed the placement choice is regime-dependent and built a budget-aware \adaptive router that traces the upper accuracy envelope of both placements on VOC while saving compute and latency within its skipping band.

Several directions remain. A \textbf{real edge-cloud testbed} would replace our offline simulation with measured round-trips~\cite{geng2025poster} on real-world hardwares. \textbf{Testing beyond VOC} such as COCO\cite{lin2014coco} would probe how far the raw-pixel routing signal carries. Beyond mAP@0.5, stricter AP@$[.5\!:\!.95]$, class-weighted or downstream-task utility will be experimented. A \textbf{skip-then-reconsider} cascade would run the \skipping estimator first and, on frames kept local where the weak pass executes anyway, apply a \conditioned estimator to revisit the offload decision on the now-available weak features at no extra detector cost. A \textbf{head-to-head with early-exit routers} across weak-model prefix depths would chart when a shallow prefix already rivals the image-only \skipping estimator.

%% file: appendix.tex
\appendix

\section{Supplementary Evidence}
\label{sec:appendix}

All numbers below are from the same single-pass VOC evaluation (seed $42$) used in \Cref{sec:evidence}, profiled on the setup of \Cref{tab:setup}. Peak~mAP is the maximum end-to-end accuracy over the offload-budget $\rho$ sweep. mAP@0.5 is the VOC metric and AP@[.5:.95] (COCO-style; written AP$_{\text C}$ in table headers) is the stricter metric. These appendices backs and justifies three design choices made in \Cref{sec:method} (the learning target, the estimator backbone, and the claim that routing signal is recoverable from the raw image) and probe robustness to a stricter accuracy metric.

\subsection{Backbone Ablation}
\label{sec:appendix-backbone}

\begin{table}[ht!]
\centering
\footnotesize
\caption{Backbone ablation for the \skipping estimator, controlling for the learning target: both trunks use the identical OffloadBin/focal target and differ only in the backbone (VOC test, seed $42$). GFLOPs and parameters are for the estimator alone. Trunks we trained on other targets are not directly comparable and are omitted.}
\label{tab:backbone-ablation}
\setlength{\tabcolsep}{5pt}
\renewcommand{\arraystretch}{1.05}
\begin{tabular}{@{}lcccc@{}}
\toprule
Backbone (OffloadBin/focal) & GFLOPs & Par.\,(M) & $\rho_s$ & Peak mAP \\
\midrule
\best{MobileNetV2-Lite} & $0.153$ & $0.54$ & \best{$0.557$} & \best{$0.808$} \\
EfficientNet-B0-Lite    & $0.176$ & $0.85$ & $0.235$ & $0.795$ \\
\bottomrule
\end{tabular}
\end{table}

Our backbone sweep is small, so we report it only briefly. \Cref{tab:backbone-ablation} holds the learning target fixed at OffloadBin/focal and varies the trunk alone: the larger EfficientNet-B0-Lite does not improve routing over the compact MobileNetV2-Lite, with a markedly lower rank correlation and no gain in peak mAP. This is consistent with the premise of \Cref{sec:intro} that detecting difficulty needs little capacity, though a broader sweep is left to future work.

\subsection{Learning-Target Ablation}
\label{sec:appendix-target}

\begin{table*}[ht!]
\centering
\footnotesize
\caption{Learning-target ablation on the fixed MobileNetV2-Lite \skipping backbone (VOC test, seed $42$). $\rho_s$ is the Spearman correlation between the estimator score and the per-frame oracle gain (not the offload budget $\rho$), and ratio-err is the mean absolute gap between realized and requested offload fractions. Exact target definitions are in our released artifacts. Rows are sorted by $\rho_s$ within each group, best per column in \best{bold}.}
\label{tab:target-ablation}
\setlength{\tabcolsep}{5pt}
\renewcommand{\arraystretch}{1.05}
\begin{tabularx}{\textwidth}{@{}l>{\raggedright\arraybackslash}Xcccc@{}}
\toprule
Target / loss & Description & $\rho_s$ & Peak mAP & Peak AP$_{\text{C}}$ & ratio-err \\
\midrule
\multicolumn{6}{@{}l}{\textit{Classification target (focal)}} \\
\quad \best{OffloadBin} & Binary $\mathbf{1}[\Delta\!\text{AP}{>}0]$: does the cloud strictly help this frame? Focal loss handles the positive-class imbalance. & \best{$0.557$} & \best{$0.808$} & \best{$0.609$} & $0.009$ \\
\quad TopQuartile & Binary: is the frame's gain in the top quartile of $\Delta\!\text{AP}$? A rarer, harder positive class than OffloadBin. & $0.389$ & $0.800$ & $0.606$ & $0.007$ \\
\midrule
\multicolumn{6}{@{}l}{\textit{Regression target}} \\
\quad MORIC$^{\!+}$ & Signed empirical CDF of $\Delta\!\text{AP}$, split at zero onto $[-1,1]$; this is our reward (\Cref{eq:moricplus}). & $0.350$ & $0.801$ & $0.607$ & $0.018$ \\
\quad HighIoUGain & Continuous per-frame gain proxy that up-weights tightly-localised, high-IoU matches. & $0.345$ & $0.794$ & $0.607$ & $0.008$ \\
\quad MORIC$^{\!+}$ (quantile) & The MORIC$^{\!+}$ target, fit with a quantile (pinball) regression loss. & $0.313$ & $0.801$ & $0.608$ & $0.010$ \\
\quad F1Gain & Continuous per-frame proxy: the change in detection F1 when the frame is offloaded. & $0.306$ & $0.791$ & $0.605$ & $0.006$ \\
\quad MORIC$^{\!+}$ (wing) & The MORIC$^{\!+}$ target, fit with a wing regression loss. & $0.303$ & $0.799$ & $0.606$ & $0.024$ \\
\quad RescueRatio & Continuous proxy for the share of a frame's missed objects that the cloud recovers. & $0.291$ & $0.792$ & $0.605$ & $0.004$ \\
\quad WorstCaseGain & Continuous proxy that emphasises the frame's worst-case detection loss. & $0.281$ & $0.796$ & $0.607$ & $0.010$ \\
\quad SigMORIC & The MORIC$^{\!+}$ reward passed through a sigmoid squashing. & $0.273$ & $0.797$ & $0.605$ & $0.015$ \\
\quad MORIC$\star$ & The MORIC$^{\!+}$ reward under an alternative CDF reshaping. & $0.271$ & $0.797$ & $0.605$ & $0.015$ \\
\quad $\Phi$-MORIC & The MORIC$^{\!+}$ reward under a $\Phi$-based CDF reshaping. & $0.263$ & $0.797$ & $0.605$ & $0.015$ \\
\quad RescueRatio (wing) & The RescueRatio target, fit with a wing regression loss. & $0.169$ & $0.791$ & $0.605$ & $0.011$ \\
\bottomrule
\end{tabularx}
\end{table*}

\Cref{sec:skipping} reduces the long-tailed per-frame $\Delta\!\text{AP}$ to a binary \emph{OffloadBin} label trained with focal loss, rather than regressing a continuous reward. Because the backbone and the target must be chosen jointly, a chicken-and-egg dependency, \Cref{tab:target-ablation} fixes the backbone first and sweeps the target. OffloadBin wins by a wide margin ($\rho_s{=}0.557$, against $0.389$ for the next-best target and $0.169 \sim 0.350$ for the continuous-regression variants) and tops both accuracy metrics. Budget tracking is uniformly tight, so this gap reflects the target itself rather than calibration. The result confirms the \Cref{sec:skipping} argument: collapsing \emph{whether offloading helps} into a binary label decouples hardness from content and lets raw pixels suffice, whereas regressing the signed magnitude wastes capacity on the $62.6\%$ zero spike (\Cref{fig:target-dist}).

\subsection{Where Offloading Helps}
\label{sec:appendix-slice}

\begin{figure}[htbp]
\centering
\includegraphics[width=\columnwidth]{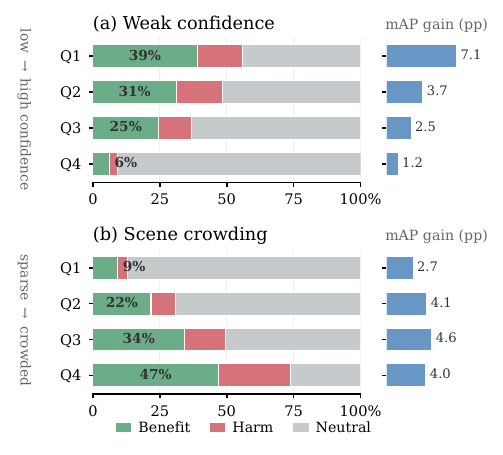}
\caption{Where offloading helps, by weak-detector stratum (VOC test, seed~$42$, $N{=}3105$). Each bar splits a quartile's frames into \textcolor{benefit}{Benefit} (offloading raises $\Delta\!\text{AP}$), \textcolor{harm}{Harm} (lowers it), and \textcolor{neutral}{Neutral} (unchanged). The dominant neutral mass shows the benefit is a sparse partition. \textbf{(a)}~Across weak-confidence quartiles benefit falls $6.4\times$ (Q1$\to$Q4). \textbf{(b)}~Across scene crowding (weak detection count) it rises $5.1\times$. The right strip is the per-frame \textcolor{headroom}{weak$\to$strong mAP gain} in points (pp), largest where benefit concentrates.}
\label{fig:slice}
\end{figure}

A \skipping estimator presumes that \emph{which frames benefit from the cloud} is both structured and recoverable from the raw image. \Cref{fig:slice} speaks to the first half: the benefit is sharply concentrated, with the lowest weak-confidence quartile offload-beneficial $6.4\times$ more often than the highest ($38.9\%$ vs.\ $6.1\%$) and the most crowded scenes $5.1\times$ more often than the sparsest ($46.9\%$ vs.\ $9.2\%$), tabulated in full in \Cref{tab:slice}. The same strata carry the largest weak$\rightarrow$strong mAP headroom. These strata are defined by weak-detector outputs, however, so these views \emph{localise} where offloading helps rather than showing the signal is recoverable \emph{before} the weak pass. We assume that low confidence and crowding are correlates of scene complexity that is plausibly legible from raw pixels and is established directly by the \skipping estimator's measured routing quality (\Cref{tab:target-ablation}).

\begin{table}[hb!]
\centering
\footnotesize
\caption{Offload benefit broken down by weak-detector stratum on the VOC test set (seed $42$, $N{=}3105$ frames). We split the frames into quartiles Q1--Q4 of each weak-detector summary, so that Q1 contains the least-confident or sparsest scenes and Q4 the most-confident or most-crowded. The Benefit, Harm, and Neutral columns give the fraction of frames in each stratum for which offloading respectively raises, lowers, or leaves $\Delta\!\text{AP}$ unchanged, and they sum to one. The final two columns, mAP$_w$ and mAP$_s$, report the mean per-frame mAP of the weak and strong detectors within the stratum.}
\label{tab:slice}
\setlength{\tabcolsep}{4pt}
\renewcommand{\arraystretch}{1.05}
\begin{tabular}{@{}lcccccc@{}}
\toprule
Stratum & Q & Benefit & Harm & Neutral & mAP$_w$ & mAP$_s$ \\
\midrule
\multirow{4}{*}{\shortstack[l]{Weak conf.\\(mean)}}
 & Q1 & $0.389$ & $0.169$ & $0.443$ & $0.750$ & $0.821$ \\
 & Q2 & $0.312$ & $0.173$ & $0.515$ & $0.847$ & $0.884$ \\
 & Q3 & $0.246$ & $0.121$ & $0.633$ & $0.895$ & $0.920$ \\
 & Q4 & $0.061$ & $0.030$ & $0.911$ & $0.961$ & $0.973$ \\
\midrule
\multirow{4}{*}{\shortstack[l]{Weak det.\\count}}
 & Q1 & $0.092$ & $0.037$ & $0.872$ & $0.928$ & $0.955$ \\
 & Q2 & $0.217$ & $0.093$ & $0.690$ & $0.864$ & $0.905$ \\
 & Q3 & $0.343$ & $0.151$ & $0.507$ & $0.831$ & $0.877$ \\
 & Q4 & $0.469$ & $0.269$ & $0.262$ & $0.784$ & $0.824$ \\
\bottomrule
\end{tabular}
\end{table}

\textbf{The benefit structure dictates the target.} \Cref{tab:target-ablation,fig:slice} are two views of one phenomenon. The offload benefit is a sparse \emph{partition}, not a smooth magnitude field: a dominant neutral mass ($62.6\%$ of frames at $\Delta\!\text{AP}{=}0$; median gain $0$ in \emph{every} stratum) around a minority of beneficial frames (${\le}47\%$ even at best) in low-confidence, crowded scenes. The learnable question is thus \emph{whether} a frame lies in that region, not \emph{by how much} it gains, which is why \Cref{tab:target-ablation} ranks both classification targets (OffloadBin $\rho_s{=}0.557$, TopQuartile $0.389$) above every regression variant ($\rho_s{\le}0.350$), and why routing reduces to recognising a region of image space that a cheap raw-image trunk can read.

%% file: paper.bbl

\begin{thebibliography}{28}


\ifx \showCODEN    \undefined \def \showCODEN     #1{\unskip}     \fi
\ifx \showISBNx    \undefined \def \showISBNx     #1{\unskip}     \fi
\ifx \showISBNxiii \undefined \def \showISBNxiii  #1{\unskip}     \fi
\ifx \showISSN     \undefined \def \showISSN      #1{\unskip}     \fi
\ifx \showLCCN     \undefined \def \showLCCN      #1{\unskip}     \fi
\ifx \shownote     \undefined \def \shownote      #1{#1}          \fi
\ifx \showarticletitle \undefined \def \showarticletitle #1{#1}   \fi
\ifx \showURL      \undefined \def \showURL       {\relax}        \fi
\providecommand\bibfield[2]{#2}
\providecommand\bibinfo[2]{#2}
\providecommand\natexlab[1]{#1}
\providecommand\showeprint[2][]{arXiv:#2}

\bibitem[Cai and Vasconcelos(2018)]%
        {cai2018cascade}
\bibfield{author}{\bibinfo{person}{Zhaowei Cai} {and} \bibinfo{person}{Nuno Vasconcelos}.} \bibinfo{year}{2018}\natexlab{}.
\newblock \showarticletitle{Cascade r-cnn: Delving into high quality object detection}. In \bibinfo{booktitle}{\emph{Proceedings of the IEEE conference on computer vision and pattern recognition}}. \bibinfo{pages}{6154--6162}.
\newblock


\bibitem[Canel et~al\mbox{.}(2019)]%
        {MLSYS2019_6bcfac82}
\bibfield{author}{\bibinfo{person}{Christopher Canel}, \bibinfo{person}{Thomas Kim}, \bibinfo{person}{Giulio Zhou}, \bibinfo{person}{Conglong Li}, \bibinfo{person}{Hyeontaek Lim}, \bibinfo{person}{David~G Andersen}, \bibinfo{person}{Michael Kaminsky}, {and} \bibinfo{person}{Subramanya Dulloor}.} \bibinfo{year}{2019}\natexlab{}.
\newblock \showarticletitle{Scaling Video Analytics on Constrained Edge Nodes}. In \bibinfo{booktitle}{\emph{Proceedings of Machine Learning and Systems}}, \bibfield{editor}{\bibinfo{person}{A.~Talwalkar}, \bibinfo{person}{V.~Smith}, {and} \bibinfo{person}{M.~Zaharia}} (Eds.), Vol.~\bibinfo{volume}{1}. \bibinfo{pages}{406--417}.
\newblock
\urldef\tempurl%
\url{https://proceedings.mlsys.org/paper_files/paper/2019/file/6bcfac823d40046dca25ef6d6d59cc3f-Paper.pdf}
\showURL{%
\tempurl}


\bibitem[Cao et~al\mbox{.}(2023)]%
        {cao2023edge}
\bibfield{author}{\bibinfo{person}{Zhiqiang Cao}, \bibinfo{person}{Zhijun Li}, \bibinfo{person}{Yongrui Chen}, \bibinfo{person}{Heng Pan}, \bibinfo{person}{Youbing Hu}, {and} \bibinfo{person}{Jie Liu}.} \bibinfo{year}{2023}\natexlab{}.
\newblock \showarticletitle{Edge-cloud collaborated object detection via difficult-case discriminator}. In \bibinfo{booktitle}{\emph{2023 IEEE 43rd International Conference on Distributed Computing Systems (ICDCS)}}. IEEE, \bibinfo{pages}{259--270}.
\newblock


\bibitem[Chen and Guestrin(2016)]%
        {chen2016xgboost}
\bibfield{author}{\bibinfo{person}{Tianqi Chen} {and} \bibinfo{person}{Carlos Guestrin}.} \bibinfo{year}{2016}\natexlab{}.
\newblock \showarticletitle{Xgboost: A scalable tree boosting system}. In \bibinfo{booktitle}{\emph{Proceedings of the 22nd acm sigkdd international conference on knowledge discovery and data mining}}. \bibinfo{pages}{785--794}.
\newblock


\bibitem[Chen et~al\mbox{.}(2015)]%
        {chen2015glimpse}
\bibfield{author}{\bibinfo{person}{Tiffany Yu-Han Chen}, \bibinfo{person}{Lenin Ravindranath}, \bibinfo{person}{Shuo Deng}, \bibinfo{person}{Paramvir Bahl}, {and} \bibinfo{person}{Hari Balakrishnan}.} \bibinfo{year}{2015}\natexlab{}.
\newblock \showarticletitle{Glimpse: Continuous, real-time object recognition on mobile devices}. In \bibinfo{booktitle}{\emph{Proceedings of the 13th ACM conference on embedded networked sensor systems}}. \bibinfo{pages}{155--168}.
\newblock


\bibitem[Everingham et~al\mbox{.}(2010)]%
        {everingham2010pascal}
\bibfield{author}{\bibinfo{person}{Mark Everingham}, \bibinfo{person}{Luc Van~Gool}, \bibinfo{person}{Christopher~KI Williams}, \bibinfo{person}{John Winn}, {and} \bibinfo{person}{Andrew Zisserman}.} \bibinfo{year}{2010}\natexlab{}.
\newblock \showarticletitle{The pascal visual object classes (voc) challenge}.
\newblock \bibinfo{journal}{\emph{International journal of computer vision}} \bibinfo{volume}{88}, \bibinfo{number}{2} (\bibinfo{year}{2010}), \bibinfo{pages}{303--338}.
\newblock


\bibitem[Geng et~al\mbox{.}(2025)]%
        {geng2025poster}
\bibfield{author}{\bibinfo{person}{Wei Geng}, \bibinfo{person}{Oguz~Kagan Altas}, \bibinfo{person}{David Guzman}, \bibinfo{person}{Giovanni Bartolomeo}, \bibinfo{person}{Nitinder Mohan}, {and} \bibinfo{person}{Joerg Ott}.} \bibinfo{year}{2025}\natexlab{}.
\newblock \showarticletitle{Poster: KUT: Towards Lightweight On-path Network Assessment for Edge Orchestration}. In \bibinfo{booktitle}{\emph{Proceedings of the 21st International Conference on emerging Networking EXperiments and Technologies}}. \bibinfo{pages}{9--11}.
\newblock


\bibitem[Geng et~al\mbox{.}(2026)]%
        {geng2026smooth}
\bibfield{author}{\bibinfo{person}{Wei Geng}, \bibinfo{person}{Xiang Su}, \bibinfo{person}{Nitinder Mohan}, \bibinfo{person}{J{\"o}rg Ott}, {and} \bibinfo{person}{Pan Hui}.} \bibinfo{year}{2026}\natexlab{}.
\newblock \showarticletitle{{SMOOTH}: Scalable Multitask Offloading with Backbone Sharing}. In \bibinfo{booktitle}{\emph{2026 IFIP Networking Conference (IFIP Networking)}}. IFIP, \bibinfo{pages}{1--10}.
\newblock


\bibitem[Guo et~al\mbox{.}(2017)]%
        {guo2017calibration}
\bibfield{author}{\bibinfo{person}{Chuan Guo}, \bibinfo{person}{Geoff Pleiss}, \bibinfo{person}{Yu Sun}, {and} \bibinfo{person}{Kilian~Q Weinberger}.} \bibinfo{year}{2017}\natexlab{}.
\newblock \showarticletitle{On calibration of modern neural networks}. In \bibinfo{booktitle}{\emph{International conference on machine learning}}. PMLR, \bibinfo{pages}{1321--1330}.
\newblock


\bibitem[He et~al\mbox{.}(2016)]%
        {he2016deep}
\bibfield{author}{\bibinfo{person}{Kaiming He}, \bibinfo{person}{Xiangyu Zhang}, \bibinfo{person}{Shaoqing Ren}, {and} \bibinfo{person}{Jian Sun}.} \bibinfo{year}{2016}\natexlab{}.
\newblock \showarticletitle{Deep residual learning for image recognition}. In \bibinfo{booktitle}{\emph{Proceedings of the IEEE conference on computer vision and pattern recognition}}. \bibinfo{pages}{770--778}.
\newblock


\bibitem[Howard et~al\mbox{.}(2019)]%
        {howard2019mobilenetv3}
\bibfield{author}{\bibinfo{person}{Andrew Howard}, \bibinfo{person}{Mark Sandler}, \bibinfo{person}{Bo Chen}, {et~al\mbox{.}}} \bibinfo{year}{2019}\natexlab{}.
\newblock \showarticletitle{Searching for {MobileNetV3}}. In \bibinfo{booktitle}{\emph{Proceedings of the IEEE/CVF International Conference on Computer Vision (ICCV)}}.
\newblock


\bibitem[Huang et~al\mbox{.}(2017)]%
        {huang2017multi}
\bibfield{author}{\bibinfo{person}{Gao Huang}, \bibinfo{person}{Danlu Chen}, \bibinfo{person}{Tianhong Li}, \bibinfo{person}{Felix Wu}, \bibinfo{person}{Laurens Van Der~Maaten}, {and} \bibinfo{person}{Kilian~Q Weinberger}.} \bibinfo{year}{2017}\natexlab{}.
\newblock \showarticletitle{Multi-scale dense networks for resource efficient image classification}.
\newblock \bibinfo{journal}{\emph{arXiv preprint arXiv:1703.09844}} (\bibinfo{year}{2017}).
\newblock


\bibitem[Jiang et~al\mbox{.}(2018)]%
        {jiang2018chameleon}
\bibfield{author}{\bibinfo{person}{Junchen Jiang}, \bibinfo{person}{Ganesh Ananthanarayanan}, \bibinfo{person}{Peter Bodik}, \bibinfo{person}{Siddhartha Sen}, {and} \bibinfo{person}{Ion Stoica}.} \bibinfo{year}{2018}\natexlab{}.
\newblock \showarticletitle{Chameleon: scalable adaptation of video analytics}. In \bibinfo{booktitle}{\emph{Proceedings of the 2018 conference of the ACM special interest group on data communication}}. \bibinfo{pages}{253--266}.
\newblock


\bibitem[Kang et~al\mbox{.}(2017)]%
        {kang2017neurosurgeon}
\bibfield{author}{\bibinfo{person}{Yiping Kang}, \bibinfo{person}{Johann Hauswald}, \bibinfo{person}{Cao Gao}, \bibinfo{person}{Austin Rovinski}, \bibinfo{person}{Trevor Mudge}, \bibinfo{person}{Jason Mars}, {and} \bibinfo{person}{Lingjia Tang}.} \bibinfo{year}{2017}\natexlab{}.
\newblock \showarticletitle{Neurosurgeon: Collaborative intelligence between the cloud and mobile edge}.
\newblock \bibinfo{journal}{\emph{ACM SIGARCH Computer Architecture News}} \bibinfo{volume}{45}, \bibinfo{number}{1} (\bibinfo{year}{2017}), \bibinfo{pages}{615--629}.
\newblock


\bibitem[Kaya et~al\mbox{.}(2019)]%
        {kaya2019shallow}
\bibfield{author}{\bibinfo{person}{Yigitcan Kaya}, \bibinfo{person}{Sanghyun Hong}, {and} \bibinfo{person}{Tudor Dumitras}.} \bibinfo{year}{2019}\natexlab{}.
\newblock \showarticletitle{Shallow-deep networks: Understanding and mitigating network overthinking}. In \bibinfo{booktitle}{\emph{International conference on machine learning}}. PMLR, \bibinfo{pages}{3301--3310}.
\newblock


\bibitem[Laskaridis et~al\mbox{.}(2020)]%
        {laskaridis2020spinn}
\bibfield{author}{\bibinfo{person}{Stefanos Laskaridis}, \bibinfo{person}{Stylianos~I Venieris}, \bibinfo{person}{Mario Almeida}, \bibinfo{person}{Ilias Leontiadis}, {and} \bibinfo{person}{Nicholas~D Lane}.} \bibinfo{year}{2020}\natexlab{}.
\newblock \showarticletitle{SPINN: Synergistic progressive inference of neural networks over device and cloud}. In \bibinfo{booktitle}{\emph{Proceedings of the 26th annual international conference on mobile computing and networking}}. \bibinfo{pages}{1--15}.
\newblock


\bibitem[Li et~al\mbox{.}(2018)]%
        {li2018edge}
\bibfield{author}{\bibinfo{person}{En Li}, \bibinfo{person}{Zhi Zhou}, {and} \bibinfo{person}{Xu Chen}.} \bibinfo{year}{2018}\natexlab{}.
\newblock \showarticletitle{Edge intelligence: On-demand deep learning model co-inference with device-edge synergy}. In \bibinfo{booktitle}{\emph{Proceedings of the 2018 workshop on mobile edge communications}}. \bibinfo{pages}{31--36}.
\newblock


\bibitem[Li et~al\mbox{.}(2020)]%
        {li2020reducto}
\bibfield{author}{\bibinfo{person}{Yuanqi Li}, \bibinfo{person}{Arthi Padmanabhan}, \bibinfo{person}{Pengzhan Zhao}, \bibinfo{person}{Yufei Wang}, \bibinfo{person}{Guoqing~Harry Xu}, {and} \bibinfo{person}{Ravi Netravali}.} \bibinfo{year}{2020}\natexlab{}.
\newblock \showarticletitle{Reducto: On-camera filtering for resource-efficient real-time video analytics}. In \bibinfo{booktitle}{\emph{Proceedings of the Annual conference of the ACM Special Interest Group on Data Communication on the applications, technologies, architectures, and protocols for computer communication}}. \bibinfo{pages}{359--376}.
\newblock


\bibitem[Lin et~al\mbox{.}(2017a)]%
        {lin2017feature}
\bibfield{author}{\bibinfo{person}{Tsung-Yi Lin}, \bibinfo{person}{Piotr Doll{\'a}r}, \bibinfo{person}{Ross Girshick}, \bibinfo{person}{Kaiming He}, \bibinfo{person}{Bharath Hariharan}, {and} \bibinfo{person}{Serge Belongie}.} \bibinfo{year}{2017}\natexlab{a}.
\newblock \showarticletitle{Feature pyramid networks for object detection}. In \bibinfo{booktitle}{\emph{Proceedings of the IEEE conference on computer vision and pattern recognition}}. \bibinfo{pages}{2117--2125}.
\newblock


\bibitem[Lin et~al\mbox{.}(2017b)]%
        {lin2017focal}
\bibfield{author}{\bibinfo{person}{Tsung-Yi Lin}, \bibinfo{person}{Priya Goyal}, \bibinfo{person}{Ross Girshick}, \bibinfo{person}{Kaiming He}, {and} \bibinfo{person}{Piotr Doll{\'a}r}.} \bibinfo{year}{2017}\natexlab{b}.
\newblock \showarticletitle{Focal loss for dense object detection}. In \bibinfo{booktitle}{\emph{Proceedings of the IEEE international conference on computer vision}}. \bibinfo{pages}{2980--2988}.
\newblock


\bibitem[Lin et~al\mbox{.}(2014)]%
        {lin2014coco}
\bibfield{author}{\bibinfo{person}{Tsung-Yi Lin}, \bibinfo{person}{Michael Maire}, \bibinfo{person}{Serge Belongie}, \bibinfo{person}{James Hays}, \bibinfo{person}{Pietro Perona}, \bibinfo{person}{Deva Ramanan}, \bibinfo{person}{Piotr Doll{\'a}r}, {and} \bibinfo{person}{C.~Lawrence Zitnick}.} \bibinfo{year}{2014}\natexlab{}.
\newblock \showarticletitle{Microsoft {COCO}: Common Objects in Context}. In \bibinfo{booktitle}{\emph{Proceedings of the European Conference on Computer Vision (ECCV)}}.
\newblock


\bibitem[Matsubara et~al\mbox{.}(2022)]%
        {matsubara2022split}
\bibfield{author}{\bibinfo{person}{Yoshitomo Matsubara}, \bibinfo{person}{Marco Levorato}, {and} \bibinfo{person}{Francesco Restuccia}.} \bibinfo{year}{2022}\natexlab{}.
\newblock \showarticletitle{Split computing and early exiting for deep learning applications: Survey and research challenges}.
\newblock \bibinfo{journal}{\emph{Comput. Surveys}} \bibinfo{volume}{55}, \bibinfo{number}{5} (\bibinfo{year}{2022}), \bibinfo{pages}{1--30}.
\newblock


\bibitem[Qiu et~al\mbox{.}(2024)]%
        {qiu2024optimizing}
\bibfield{author}{\bibinfo{person}{Jiaming Qiu}, \bibinfo{person}{Ruiqi Wang}, \bibinfo{person}{Brooks Hu}, \bibinfo{person}{Roch Gu{\'e}rin}, {and} \bibinfo{person}{Chenyang Lu}.} \bibinfo{year}{2024}\natexlab{}.
\newblock \showarticletitle{Optimizing edge offloading decisions for object detection}. In \bibinfo{booktitle}{\emph{2024 IEEE/ACM Symposium on Edge Computing (SEC)}}. IEEE, \bibinfo{pages}{164--177}.
\newblock


\bibitem[Ren et~al\mbox{.}(2015)]%
        {ren2015fasterrcnn}
\bibfield{author}{\bibinfo{person}{Shaoqing Ren}, \bibinfo{person}{Kaiming He}, \bibinfo{person}{Ross Girshick}, {and} \bibinfo{person}{Jian Sun}.} \bibinfo{year}{2015}\natexlab{}.
\newblock \showarticletitle{Faster {R-CNN}: Towards Real-Time Object Detection with Region Proposal Networks}. In \bibinfo{booktitle}{\emph{Advances in Neural Information Processing Systems (NeurIPS)}}.
\newblock


\bibitem[Sandler et~al\mbox{.}(2018)]%
        {sandler2018mobilenetv2}
\bibfield{author}{\bibinfo{person}{Mark Sandler}, \bibinfo{person}{Andrew Howard}, \bibinfo{person}{Menglong Zhu}, \bibinfo{person}{Andrey Zhmoginov}, {and} \bibinfo{person}{Liang-Chieh Chen}.} \bibinfo{year}{2018}\natexlab{}.
\newblock \showarticletitle{Mobilenetv2: Inverted residuals and linear bottlenecks}. In \bibinfo{booktitle}{\emph{Proceedings of the IEEE conference on computer vision and pattern recognition}}. \bibinfo{pages}{4510--4520}.
\newblock


\bibitem[Teerapittayanon et~al\mbox{.}(2016)]%
        {teerapittayanon2016branchynet}
\bibfield{author}{\bibinfo{person}{Surat Teerapittayanon}, \bibinfo{person}{Bradley McDanel}, {and} \bibinfo{person}{Hsiang-Tsung Kung}.} \bibinfo{year}{2016}\natexlab{}.
\newblock \showarticletitle{Branchynet: Fast inference via early exiting from deep neural networks}. In \bibinfo{booktitle}{\emph{2016 23rd international conference on pattern recognition (ICPR)}}. IEEE, \bibinfo{pages}{2464--2469}.
\newblock


\bibitem[Viola and Jones(2001)]%
        {viola2001rapid}
\bibfield{author}{\bibinfo{person}{Paul Viola} {and} \bibinfo{person}{Michael Jones}.} \bibinfo{year}{2001}\natexlab{}.
\newblock \showarticletitle{Rapid object detection using a boosted cascade of simple features}. In \bibinfo{booktitle}{\emph{Proceedings of the 2001 IEEE computer society conference on computer vision and pattern recognition. CVPR 2001}}, Vol.~\bibinfo{volume}{1}. Ieee, \bibinfo{pages}{I--I}.
\newblock


\bibitem[Wang et~al\mbox{.}(2024)]%
        {wang2024tiny}
\bibfield{author}{\bibinfo{person}{Qingyuan Wang}, \bibinfo{person}{Barry Cardiff}, \bibinfo{person}{Antoine Frapp{\'e}}, \bibinfo{person}{Benoit Larras}, {and} \bibinfo{person}{Deepu John}.} \bibinfo{year}{2024}\natexlab{}.
\newblock \showarticletitle{Tiny models are the computational saver for large models}. In \bibinfo{booktitle}{\emph{European Conference on Computer Vision}}. Springer, \bibinfo{pages}{163--182}.
\newblock


\end{thebibliography}
